\documentclass[twocolumn,showpacs,preprintnumbers,amsmath,amssymb]{revtex4}
\usepackage{graphicx}
\usepackage{dcolumn}
\usepackage{bm,latexsym}

\def\Journal#1#2#3#4{{#1} {\bf #2}, #3 (#4)}

\def\PLB{{Phys. Lett.}  B}
\def\PRL{Phys. Rev. Lett.}

\def\PRD{{Phys. Rev.} D}

\def\APJ{{Astrophys. J.}}

\def\NAT{{Nature}}

\def\CQG{{Class. Quant. Grav.}}

\def\RMF{{Rev. Mex. F\'\i s.}}

\def\be{\begin{equation}}
\def\ee{\end{equation}}
\def\bea{\begin{eqnarray}}
\def\eea{\end{eqnarray}}

\begin{document}
\preprint{APS/123-QED}

\title{On the linear and weak-field limits of scalar-tensor theories of gravity}
\author{ Marcelo Salgado} 
\email{marcelo@nuclecu.unam.mx}
\homepage{http://www.nuclecu.unam.mx/~marcelo}
\affiliation{Instituto de Ciencias Nucleares 
\\ Universidad Nacional Aut\'onoma de M\'exico 
\\ Apdo. Postal 70--543 M\'exico 04510 D.F., M\'exico}

\date{\today}

\begin{abstract}
The linear approximation of scalar-tensor theories of gravity is
obtained in the physical (Jordan) frame under the 4+0 (covariant) and 3+1
formalisms. Then the weak-field limit is analyzed and the conditions leading
to significant deviations of the $1/r^{2}$ Newton's law of gravitation are discussed.
Finally, the scalarization effects induced by these theories in extended
objects are confronted within the weak-field limit.
\end{abstract}

\pacs{04.50.+h,04.25.Nx,04.80.Cc}

\maketitle

\section{Introduction}

In the last decade there has been an increasing interest in the so-called
scalar-tensor theories of gravity (STT; see Ref. \cite{damour1} for a
review) in view of the possible deviations that Einstein's general theory of
relativity (GR) could show in the framework of several upcoming
observations. Among these, there has been a considerable interest in
detecting a scalar-wave component (spin-0 waves) in addition to the ordinary
gravitational waves (spin-2 waves) predicted by GR (see for instance 
\cite{wagoner2,wagoner1,brunetti,barbusci,nakao}). Among the potential emitters of such
scalar-waves are the compact binary systems and neutron stars. Yet, the
binary pulsar has showed no appreciable deviations from GR due to the
emission of scalar waves, and then one is only able to put stringent bound
in the STT parameters or couplings \cite{damour2}. Moreover, through the
effects of spontaneous scalarization in neutron stars 
\cite{damour2,damour3,novak, salgado2}, it seems that
scalar-waves have also a small chance to be observed in detectors like VIRGO
or LIGO within a few hundreds of kiloparsecs \cite{novak}. However, there is
the hope that resonant mass detectors of spherical shape or interferometers
like LISA can resolve the existence of a scalar-gravitational wave 
\cite{brunetti,barbusci,nakao}.

At the large scale (cosmological scales), STT have been proposed as models
for dark energy that can replace the cosmological constant 
\cite{dark}, and also for
explaining some other features of the galaxy distribution in our universe 
\cite{us}. 
However, detailed analysis show that when all the observational constraints
(both cosmological and local) are taken into account, the simpler STT fail
or are extremely constrained \cite{salgado}. At this point one cannot only
but recognizes the predictive power of GR, confirmed by experiments and
observations that were maybe far from being imagined by Einstein at the time
of the creation of his theory. At smaller scales (scales of the order of
meters or kilometers) the spurious discovery of a fifth force renewed the
idea about the existence of new fundamental fields of meter-range \cite
{fishbach}. At this regard, considerable effort has been put in measuring
gravitationally such kind of interactions that can mimic a varying
gravitational ``constant'' $G$.

In this report the linear limit of STT is reviewed under the usual 4+0
formalism. The weak-field limit is taken and
confronted in view of the past and recent experiments (ground based and
satellite mission) intended to test the gravitational interaction between
bodies. The analysis is performed in the Jordan frame, and therefore no
intermediate unphysical variables or transformations are introduced. A
Yukawa potential appears as a new term in addition to the ordinary Newtonian
potential due to the non-minimal coupling between the scalar field with
gravity, the entire gravitational potential having been identified with the
potential of test particles. It is then argued that the Yukawa potential can
produce significative deviations on the $1/r^{2}$ Newton's gravitation law
if the scalar ``particles'' are massive enough.

An appendix analyzing the linear limit of STT under the 3+1 formulation of
GR is to be found at the end.

\section{Scalar-Tensor Theories of Gravity}

The general action for a scalar-tensor theory of gravity with a single
scalar field is given by 
\begin{eqnarray}
  \label{jordan}
S[g_{\mu\nu}, \phi, \psi] &=& \int \left\{ \frac{1}{16\pi G_0} F(\phi) R
-\left( \frac{1}{2}(\nabla \phi)^2 + V(\phi) \right) \right\} 
\nonumber \\
&&\times  \sqrt{-g} d^4x + S[g_{\mu\nu}, \psi]
\end{eqnarray}
where $\psi$ represents collectively the matter fields (fields other than $%
\phi$).

The representation of the scalar-tensor theories given by Eq. (\ref{jordan})
is called the {\it Jordan frame} representation. One can parametrize the
same theories as 
\begin{eqnarray}
  \label{BD}
S[g_{\mu\nu}, \Phi, \psi] &=& \frac{1}{16\pi G_0} \int \left\{ \Phi R - \frac{%
\omega(\Phi)}{\Phi} (\nabla \Phi)^2 + 2\Phi \lambda (\Phi) \right\}\nonumber 
\\ && \times \sqrt{-g} d^4 x + S[g_{\mu\nu}, \psi]
\end{eqnarray}
where 
\begin{eqnarray}  \label{BDpar}
\Phi &:=& F(\phi) \,\,\,\,, \\
\omega_{{\rm BD}} (\Phi) &:=& \frac{8\pi G_0\Phi}{(\partial_\phi F)^2}
\,\,\,\,. \\
\lambda (\Phi) &:=& -\frac{8\pi V(\phi)}{\Phi} \,\,\,\,.
\end{eqnarray}
For instance, the Jordan-frame representation of the Brans-Dicke theory with 
$\omega=$const. corresponds to $F= 2\pi G_0 \phi^2/\omega$ and $V(\phi) =0$.
It is also customary to parametrize the scalar-tensor theories in the
so-called {\it Einstein frame} by introducing {\it non-physical} fields as
follows, 
\begin{eqnarray}
g_{\mu\nu}^* &:=& F(\phi) g_{\mu\nu} \,\,\,\,, \\
\phi^* &=& \int \left[ \frac{3}{4}\frac{1}{F^2(\phi)}\left( \partial_\phi
F\right)^2 + \frac{4\pi}{F(\phi)}\right]^{1/2} d\phi \,\,\,\,, \\
F^*(\phi^*) &=& F(\phi) \,\,\,\,.
\end{eqnarray}
so that the action Eq. (\ref{jordan}) takes the form 
\begin{eqnarray}
 \label{einst}
S[g^*_{\mu\nu}, \phi^*, \psi] &=& \frac{1}{16\pi G_0} \int \left[ R^* - 2
(\nabla^* \phi^*)^2 - V^*(\phi^*) \right] \nonumber \\
&& \times \sqrt{-g^*} d^4 x +
S[g_{\mu\nu}^*/F^*(\phi^*), \psi] \,\,\,\,,
\end{eqnarray}
where all quantities with `*' are computed with the non-physical metric $%
g_{\mu\nu}^*$ and $\phi^*$.

We emphasize that although the equations of motions obtained from the
Einstein frame are simpler than those from the Jordan frame, in the sense
that the field $\phi ^{*}$ appears to be coupled minimally to the
non-physical metric, the matter equations derived from the Bianchi
identities $\nabla _{\mu }^{*}G_{\mu \nu }^{*}=0$ will have sources, i.e., $%
\nabla _{\mu }^{*}T_{\psi }^{*\,\,\mu \nu }\neq 0$, where here $T_{\psi
}^{*\,\,\mu \nu }=T_{\psi }^{\mu \nu }/{F^{*}}^{2}(\phi ^{*})$ is the non
physical energy-momentum tensor of the matter fields $\psi $. However, in
the Jordan frame the matter equations resulting from the Bianchi identities $%
\nabla _{\mu }G^{\mu \nu }=0$ turn to satisfy $\nabla _{\mu }T_{\psi }^{\mu
\nu }=0$, reflecting explicitly the fulfillment of the Einstein's weak 
equivalence principle (that is the origin of the name ``physical metric'').

In the following the Jordan frame representation of the scalar-tensor
theories will only be used. The equations of motion obtained from a
variational principle using Eq. (\ref{jordan}) are
\begin{eqnarray}  
G^{\mu\nu} &=& 8\pi G_0 T^{\mu\nu}\,\,\,\,, \\
\label{effTmunu}
T^{\mu\nu} &:=& \frac{G_{{\rm eff}}}{F}\left( T^{\mu\nu}_F + T^{\mu\nu}_{%
{\rm sf}} + T^{\mu\nu}_{{\rm matt}}\right)\,\,\,\,, \\
T^{\mu\nu}_F &:= & \frac{1}{8\pi G_0}\left[\nabla^\mu\left(\partial_\phi
F\nabla^\nu\phi\right) - g^{\mu\nu}\nabla_\lambda \left(\partial_\phi
F\nabla^\lambda \phi\right)\right] \,\,\,\,,\nonumber \\
&& \\
T^{\mu\nu}_{{\rm sf}} &:= & (\nabla^\mu \phi)(\nabla^\nu \phi) - g^{\mu\nu}
\left[ \frac{1}{2}(\nabla \phi)^2 + V(\phi)\right ] \,\,\,\,, \\
\label{Geff}
G_{{\rm eff}} &:=& \frac{G_0}{F} \,\,\,\,, 
\end{eqnarray}
\begin{widetext}
\begin{equation}
\label{KG}
{\Box \phi} = \frac{ F\partial_\phi V- 2(\partial_\phi F) V -\frac{1}{2}
(\partial_\phi F) \left( 1 + \frac{3\partial^2_{\phi\phi} F}{8\pi G_0}
\right)(\nabla \phi)^2 + \frac{1}{2}(\partial_\phi F) T_{{\rm matt}} }{ F + 
\frac{ 3(\partial_\phi F)^2}{16\pi G_0} }\,\,\,\,.
\end{equation}
\end{widetext}
where $T_{{\rm matt}}$ stands for the trace of $T^{\mu\nu}_{{\rm matt}}$ and
the subscript ``matt'' refers to the matter fields other that $\phi$.

Now, the Bianchi identities imply 
\begin{equation}
\nabla _{\mu }T^{\mu \nu }=0\,\,\,\,.
\end{equation}
However, the use of the equations of motion leads as mentioned to the
energy-conservation equations of matter 
\begin{equation}
\nabla _{\mu }T_{{\rm matt}}^{\mu \nu }=0\,\,\,\,,
\end{equation}
which implies in the case of test particles, that bodies are subject to no
other long range forces than the gravitational ones (free falling
particles). In other words, the scalar field being not directly coupled to
matter no direct interaction between the scalar field and matter
arises. The scalar
field $\phi $ will only interact with matter gravitationally, i.e., only
through the curvature effects.

The final form of the field equations have exactly the same form as in
general relativity with an effective energy-momentum tensor. This means that
one can formulate the Cauchy problem for scalar-tensor theories exactly in
the same manner as in GR (see the Appendix).

\section{Linear limit of STT}

The linear limit of STT has been analyzed in the past by many authors.
Wagoner \cite{wagoner2} was one of the first in analyzing the gravitational
wave emission in STT and the weak-field approximation (an updated analysis
was performed in \cite{wagoner1}). Such an analysis was performed in the
Einstein frame. More recently, Pimentel and Obreg\'{o}n \cite{pimentel}
performed a similar analysis in the Brans-Dicke representation as in Eq. 
(\ref{BD}).

As mentioned, the linear limit of STT treated here is performed in the
Jordan frame. As usual, we consider first order perturbations of the
Minkowski spacetime: 
\begin{eqnarray}
g_{\mu \nu } &\approx &\eta _{\mu \nu }+\epsilon \gamma _{\mu \nu }\,\,\,\,,
\\
T_{\mu \nu } &\approx &T_{\mu \nu }^{0}+\epsilon \tilde{T}_{\mu \nu
}\,\,\,\,, \\
\phi &\approx &\phi _{0}+\epsilon \tilde{\phi}\,\,\,\,, \\
F(\phi ) &\approx &F_{0}+\epsilon F_{0}^{\prime }\tilde{\phi}\,\,\,\,, \\
\partial _{\phi }F(\phi ) &\approx &F_{0}^{\prime }+\epsilon F_{0}^{\prime
\prime }\tilde{\phi}\,\,\,\,, \\
\partial _{\phi \phi }^{2}F(\phi ) &\approx &F_{0}^{\prime \prime }+\epsilon
F_{0}^{\prime \prime \prime }\tilde{\phi}\,\,\,\,, \\
V(\phi ) &\approx &V_{0}+\epsilon V_{0}^{\prime }\tilde{\phi}\,\,\,\,, \\
\partial _{\phi }V(\phi ) &\approx &V_{0}^{\prime }+\epsilon V_{0}^{\prime
\prime }\tilde{\phi}\,\,\,\,.
\end{eqnarray}
where $\epsilon \ll 1$ and the knott indicates quantities at zero order. In
the 4+0 covariant formulation one can introduce the combination 
\begin{eqnarray}
\tilde{\gamma}_{\mu \nu }:= &&\bar{\gamma}_{\mu \nu }+\kappa \eta _{\mu \nu }
\tilde{\phi}\,\,\,\,,  \label{tilgamma} \\
\label{bargamma}
\bar{\gamma}_{\mu \nu }:= &&\gamma _{\mu \nu }-\frac{1}{2}\eta _{\mu \nu
}\gamma \,\,\,\,,
\end{eqnarray}
where $\gamma =\gamma _{\,\,\,\mu }^{\mu }$ and $\kappa $ is a gauge
constant to be fixed later in order to simplify the equations. The resulting
linearized Einstein equations are 
\begin{eqnarray}
\tilde{G}_{\mu \nu }&=& \partial ^{\sigma }\partial _{(\nu }\tilde{\gamma}_{\mu
)\sigma }-\frac{1}{2}\Box _{\eta }\tilde{\gamma}_{\mu \nu }-\frac{1}{2}\eta
_{\mu \nu }\partial ^{\sigma }\partial ^{\alpha }\tilde{\gamma}_{\sigma
\alpha }\nonumber \\
&=& 8\pi G_{0}\tilde{T}_{\mu \nu }+\kappa \left( \partial _{\mu \nu
}^{2}\tilde{\phi}-\eta _{\mu \nu }\Box _{\eta }\tilde{\phi}\right) \,\,\,\,.
\label{einstlin}
\end{eqnarray}
where $T_{\mu \nu }^{0}=0$ results from the self-consistency of the
perturbations at first order. Here $\Box _{\eta }$ is the D'Alambertian
operator compatible with the flat metric $\eta _{\mu \nu }$. Moreover, the
flat metric is used to raise and lower indices of first order tensorial
quantities.

The Lorentz gauge 
\begin{equation}  \label{radgaugescal0}
\partial^\mu \tilde \gamma_{\mu\nu} = 0 \,\,\,\,,
\end{equation}
generalizes the usual Lorentz gauge of GR, and can be imposed to simplify
the equations. Then, from Eq. (\ref{einstlin}) the resulting wave equation is 
\begin{equation}  \label{wavetilgamma}
\Box_\eta \tilde \gamma_{\mu\nu} = - 16\pi G_0 \tilde T_{\mu\nu} - 2\kappa
\left( \partial^2_{\mu\nu}\tilde\phi -\eta_{\mu\nu}\Box_\eta \tilde\phi
\right) \,\,\,\,.
\end{equation}

The linear approximation of the effective energy-momentum tensor Eq. (\ref
{effTmunu}) and the Klein-Gordon Eq. (\ref{KG}) turn to be 
\begin{eqnarray}\label{Tmunulin}
& & \tilde T_{\mu\nu} = \frac{\tilde T_{\mu\nu}^{{\rm matt}}}{F_0} + 
\frac{F^{\prime}_0}{8 \pi F_0 G_0} \left( \partial^2_{\mu\nu}\tilde\phi
-\eta_{\mu\nu}\Box_\eta \tilde\phi \right) \,\,\,, \\
&& \Box_\eta \tilde \phi - m_0^2 \tilde \phi = 4\pi \alpha 
\frac{F^{\prime}_0}{F_0} \tilde T_{{\rm matt}} \,\,\,\,, \\
\label{msq}
&& m_0^2 := \frac{V^{\prime\prime}_0}{1 + \frac{ 3(F^{\prime}_0)^2}{16\pi
F_0 G_0}} \,\,\,\,, \\
\label{alp}
&& \alpha := \frac{1}{8\pi \left(1 + \frac{ 3(F^{\prime}_0)^2}{16\pi F_0 G_0}\right)} 
\,\,\,\,,
\end{eqnarray}
where we used the following conditions 
\begin{equation}  \label{V0cond}
V(\phi_0) = 0 = V^{\prime}_0 = 0 \,\,\,\,,
\end{equation}
resulting from the consistency at first order of the linearized Einstein and
Klein-Gordon equations and assuming that $T_{{\rm matt}\,\,\,\mu\nu}^0 = 0$.
In this way the wave equation Eq. (\ref{wavetilgamma}) becomes 
\begin{equation}  \label{wavetilgamma2}
\Box_\eta \tilde \gamma_{\mu\nu} = - \frac{16\pi G_0}{F_0} 
\tilde T_{\mu\nu}^{{\rm matt}} - 2\left(\kappa + \frac{F^{\prime}_0}{F_0}\right)
\left( \partial^2_{\mu\nu}\tilde\phi -\eta_{\mu\nu}\Box_\eta \tilde\phi
\right) \,\,\,\,.
\end{equation}
Note that at this order the application of the ordinary divergence in Eq. 
(\ref{wavetilgamma2}) and the use of the Lorentz gauge Eq. (\ref
{radgaugescal0}) leads to the energy-conservation of the matter
perturbations: $\partial^\mu \tilde T_{\mu\nu}^{{\rm matt}}=0$.

The choice 
\begin{equation}
\kappa =-\frac{F_{0}^{\prime }}{F_{0}}\,\,\,\,,  \label{kappa}
\end{equation}
for the gauge constant simplifies the Eq. (\ref{wavetilgamma2})
considerably. So summarizing, we have the following wave equations for the
gravitational and scalar modes: 
\begin{eqnarray}
&&\Box _{\eta }\tilde{\gamma}_{\mu \nu }=-\frac{16\pi G_{0}}{F_{0}}\tilde{T}_{\mu \nu }^{{\rm matt}}\,\,\,\,,  \label{wavetilgamma3} \\
&&\partial ^{\mu }\tilde{\gamma}_{\mu \nu }=0\,\,\,\,, \\
&&\Box _{\eta }\tilde{\phi}-m_{0}^{2}\tilde{\phi}=4\pi \alpha \frac{F_{0}^{\prime }}{F_{0}}\tilde{T}_{{\rm matt}}\,\,\,\,,
\end{eqnarray}
with the constants $m_{0}^{2},\alpha$ and $\kappa$ given by Eqs. 
(\ref{msq}), (\ref{alp}) and (\ref{kappa}) respecteively. 
The analysis of propagation of gravitational
and scalar waves will be not pursued here, this has been done elsewhere 
(see Refs. \cite{barbusci,brunetti,nakao,wagoner2,wagoner1} ).

\subsection{The weak-field approximation}

In the weak-field approximation one considers slow varying fields and
sources such that the temporal derivatives in the D'Alambertian can be
neglected and $T_{\mu \nu }^{{\rm matt}}\approx \tilde{\rho}\delta _{\mu
}^{0}\delta _{\nu }^{0}$. The wave equations read then, 
\begin{eqnarray}
&&\,^{3}\tilde{\Delta}\tilde{\gamma}_{\mu \nu }=-\frac{16\pi G_{0}}{F_{0}}\tilde{T}_{\mu \nu }^{{\rm matt}}\,\,\,\,, \\
&&\,^{3}\tilde{\Delta}\tilde{\phi}-m_{0}^{2}\tilde{\phi}=4\pi \alpha \frac{F_{0}^{\prime }}{F_{0}}\tilde{T}_{{\rm matt}}\,\,\,\,,
\end{eqnarray}
where $\,^{3}\tilde{\Delta}$ is the three dimensional Euclidean Laplacian.
The condition for slow varying sources $\tilde{T}_{i0}^{{\rm matt}}\approx
0\approx \tilde{T}_{ij}^{{\rm matt}}$, leads then to the following equations 
\begin{eqnarray}
 \label{laplacetilgamma}
&&\,^{3}\tilde{\Delta}\tilde{\gamma}_{00}=-\frac{16\pi G_{0}}{F_{0}}\tilde{\rho}\,\,\,\,,  \\
 \label{laplacephi}
&&\,^{3}\tilde{\Delta}\phi -m_{0}^{2}\tilde{\phi}=-4\pi \alpha \frac{F_{0}^{\prime }}{F_{0}}\tilde{\rho}\,\,\,\,, \\
&&\tilde{\gamma}_{0i}={\rm const.}=\tilde{\gamma}_{ij}\,\,\,\,,
\end{eqnarray}
where $\tilde{\rho}:=\tilde{T}_{00}^{{\rm matt}}=-\tilde{T}^{{\rm matt}}$.
The constant solutions for $\tilde{\gamma}_{\mu i}$ arise by demanding a
well behavior of the metric at spatial infinity (without lost of generality
the constants can be gauged out). Thus, with $\tilde{\gamma}_{ij}=0$ and 
Eqs. (\ref{tilgamma}) and (\ref{bargamma}), it yields 
\begin{equation}
\tilde{\gamma}_{\mu \nu }=\gamma _{\mu \nu }-\eta _{\mu \nu }\left( \gamma
_{00}+2\kappa \tilde{\phi}\right) \,\,\,\,,
\end{equation}
and therefore 
\begin{equation}
\tilde{\gamma}_{00}=2\left( \gamma _{00}+\kappa \tilde{\phi}\right) \,\,\,\,.
\end{equation}
Since the geodesic equation for slow particles leads to the identification
of 
\begin{equation}
\Phi _{N}=-\frac{1}{2}\gamma _{00}\,\,\,\,,  \label{Newtpot0}
\end{equation}
with the Newtonian potential \cite{wald}, it turns 
\begin{equation}
\Phi _{N}=-\frac{1}{4}\tilde{\gamma}_{00}+\frac{\kappa }{2}\tilde{\phi}\,\,\,\,.
\end{equation}
The two gravitational degrees of freedom that remain in the weak-field limit
are thus $\gamma _{00}$ and $\tilde{\phi}$ ($\gamma _{0j}=0$, and $\gamma
_{ij\text{ }}$is given in terms of $\gamma _{00}$ and $\tilde{\phi}$).

The solutions of Eqs. (\ref{laplacetilgamma}) and (\ref{laplacephi}) are 
respectively 
\begin{eqnarray}  \label{phisol}
\tilde{\psi} &=&-\frac{G_{0}}{F_{0}}\int \frac{\tilde{\rho}(\vec{x}^{\prime
})}{|\vec{x}-\vec{x}^{\prime }|}d^{3}x^{\prime
}\,\,\,\,\,\,\,\,\,+\,\,\,\,{\rm B.C.}\,\,\,\,, \\
\tilde{\phi} &=&\alpha \frac{F_{0}^{\prime }}{F_{0}}\int \frac{\tilde{\rho}(
\vec{x}^{\prime })e^{-m_{0}|\vec{x}-\vec{x}^{\prime }|}}{|\vec{x}-\vec{x}
^{\prime }|}d^{3}x^{\prime }\,\,\,\,\,\,\,\,+\,\,\,\,{\rm B.C.}
\,\,,
\end{eqnarray}
where $\tilde{\psi}:=-\frac{1}{4}\tilde{\gamma}_{00}$. So finally 
\begin{widetext}
\begin{equation}
\Phi _{N}= -\frac{G_{0}}{F_{0}}\int \frac{\tilde{\rho}(\vec{x}^{\prime })}{|
\vec{x}-\vec{x}^{\prime }|}d^{3}x^{\prime }\,\,\,\, -\frac{\alpha }{2}\left( 
\frac{F_{0}^{\prime }}{F_{0}}\right) ^{2}\int \frac{\tilde{\rho}(\vec{x}
^{\prime })e^{-m_{0}|\vec{x}-\vec{x}^{\prime }|}}{|\vec{x}-\vec{x}^{\prime }|
}d^{3}x^{\prime }\,\,\,\,\,\,+\,\,\,\,\,\,{\rm B.C.}  \label{Newtpot}
\end{equation}
\end{widetext}
It is important to mention that the details of the general solution
(interior plus exterior solutions, both matched continuously at the surface
of the extended object) depend strongly on the boundary conditions. 

\subsection{Regular solutions}

Explicit solutions of the gravitational and scalar fields depend on $\tilde{%
\rho}(\vec{x})$, and as mentioned on the boundary conditions as well. Clearly, a
detailed solution of the whole system involves also the equation of
hydrostatic equilibrium and the equation of state of matter \cite{comm} .
However, here we do not want to be so explicit and only exhibit the possible
deviations that the Newton's gravitation law can suffer depending on the
values of the parameters. For simplicity spherical symmetry is assumed. We
seek for solutions for $\tilde{\psi}$ and $\tilde{\phi}$ with regularity
conditions at $r=0$ 
\begin{equation}
\partial _{r}\tilde{\psi}=0=\partial _{r}\tilde{\phi}\,\,\,\,,
\end{equation}
and asymptotic flatness conditions 
\begin{eqnarray}
&&{\rm lim}_{r\rightarrow \infty }\tilde{\psi}\rightarrow 0\,\,\,\,, \\
&&{\rm lim}_{r\rightarrow \infty }\tilde{\phi}\rightarrow 0\,\,\,\,.
\end{eqnarray}
The solutions are then given by 
\begin{eqnarray}
\tilde{\psi}(r) &=&-\frac{G_{0}}{F_{0}}\int_{r}^{+\infty }\frac{m(\hat{r})}{{%
\hat{r}}^{2}}d\hat{r}\,\,\,\,, \\
\tilde{\phi}(r) &=&\int_{r}^{+\infty }\frac{q(\hat{r})}{{\hat{r}}^{2}}d\hat{r%
}\,\,\,\,,
\end{eqnarray}
where 
\begin{eqnarray}
m(r):= &&4\pi \int_{0}^{r}\tilde{\rho}(\hat{r}){\hat{r}}^{2}d\hat{r}\,\,\,\,,
\\
q(r):= &&4\pi \int_{0}^{r}\left( \alpha \frac{F_{0}^{\prime }}{F_{0}}\tilde{
\rho}(\hat{r})-\frac{m_{0}^{2}}{4\pi }\tilde{\phi}\right) {\hat{r}}^{2}d\hat{
r}\,\,\,\,.
\end{eqnarray}
One can define global quantities as 
\begin{eqnarray}
M &:=&{\rm lim}_{r\rightarrow \infty }\left[ \frac{r^{2}}{G_{0}}\partial _{r}
\tilde{\psi}\right] \nonumber \\
&= & 4\pi \int_{0}^{+\infty }\tilde{\rho}(\hat{r}){\hat{r}}
^{2}d\hat{r}\,\,\,\,,  \label{charge} \\
Q &:= & -{\rm lim}_{r\rightarrow \infty }\left[ r^{2}\partial _{r}\tilde{\phi}
\right] \nonumber \\ 
&=& 4\pi \int_{0}^{+\infty }\left( \alpha \frac{F_{0}^{\prime }}{F_{0}}
\tilde{\rho}(\hat{r})-\frac{m_{0}^{2}}{4\pi }\tilde{\phi}\right) {\hat{r}}
^{2}d\hat{r}\,\,\,\,.
\end{eqnarray}
Note that in the minimal-coupling case ($F_{0}^{\prime }=0$), $Q\equiv 0$,
since the only regular solution of the Helmholtz equation is $\tilde{\phi}=0$%
).

If the density $\tilde{\rho}$ has compact support (as it is usually the case
for astrophysical bodies), then it vanishes for $r\geq R$, where $R$ is the
radius of the body. The solutions for $\tilde{\psi}$ and $\tilde{\phi}$
matched continuously at $r=R$ can be then written as


\begin{eqnarray}
\tilde \psi (r) &=& \left\{ \begin{array}{c} 
-\frac{G_0M}{RF_0} - \frac{G_0}{F_0}
\int_r^R \frac{m(\hat r)}{{\hat r}^2} d\hat r \,\,\,{\rm for} \,\,\, r\leq R
\,\,\,\,, \\
 -\frac{G_0M}{F_0 r} \,\,\,\, \,\,\,\, \,\,\,\, \,\,\,\,
\,\,\,\, \,\,\,\, \,\,\,\, \,\,\,\, \,\,\,\, \,\,\,\, \,\,\,\, {\rm for}
\,\,\, r\geq R \,\,\,\,, 
\end{array}\right. 
\\
\tilde \phi (r) &=& \left\{ \begin{array}{c}
 \frac{\tilde Q}{R(1+ m_0 R)} + \int_r^R
\frac{q(\hat r)}{{\hat r}^2} d\hat r \,\,\,\, {\rm for} \,\,\, r\leq R
\,\,\,\,, \\
\frac{\tilde Q e^{-m_0 r\left(1-R/r\right)} }{r(1+ m_0 R)}
\,\,\,\, \,\,\,\, \,\,\,\, \,\,\,\, \,\,\,\, \,\,\,\, \,\,\,\, {\rm for}
\,\,\, r\geq R \,\,\,\,,
\end{array}\right.
\end{eqnarray}
where 
\begin{eqnarray}
m(r) &:=& 4\pi \int_0^{r\leq R} \rho(\hat r) {\hat r}^2 d\hat r \,\,\,\,, \\
q(r)&:=& 4\pi \int_0^{r\leq R} \left( \alpha \frac{F^{\prime}_0}{F_0} \rho(%
\hat r) - \frac{m_0^2}{4\pi} \tilde \phi \right) {\hat r}^2 d\hat r \,\,\,\,,
\end{eqnarray}
and $M := m(R) $, $\tilde Q := q(R) $.

\section{Observational constraints}

In the following we consider only the exterior solutions and confronted them
with observations. Therefore, the exterior Newtonian potential (\ref{Newtpot}%
) is given by 
\begin{equation}
\Phi _{N}=-\frac{G_{0}M}{F_{0}r}\left[ 1+\frac{\alpha \sigma}{2G_{0}}\frac{%
(F_{0}^{\prime })^{2}}{F_{0}}e^{-m_{0}r}\right]
\end{equation}
where $\sigma$ is a constant that depends on the global properties of the source.
Namely, 
\begin{eqnarray}
\sigma  &=& \frac{\tilde{Q}e^{m_{0}R}F_{0}}{(1+m_{0}R)M\alpha F_{0}^{\prime }}\nonumber \\
&= & \frac{
e^{m_{0}R}}{1+m_{0}R}\left( 1-\frac{m_{0}^{2}F_{0}}{M\alpha F_{0}^{\prime }}
\int_{0}^{R}\tilde{\phi}(r)r^{2}dr\right) \,\,\,\,.  \label{AA}
\end{eqnarray}
It is customary to express the coefficients involving $F_{0}^{\prime }$ in
terms of the effective Brans-Diecke parameter Eq.(\ref{BDpar}) 
\begin{equation}
\omega _{{\rm BD}}^{0}=\frac{8\pi G_{0}F_{0}}{(F_{0}^{\prime })^{2}}\,\,\,\,.
\end{equation}
Then, 
\begin{eqnarray}
\alpha &=&\frac{\omega _{{\rm BD}}^{0}}{4\pi \left( 3+2\omega _{{\rm BD}%
}^{0}\right) }\,\,\,\,, \\
m_{0}^{2} &=&\frac{2\omega _{{\rm BD}}^{0}V_{0}^{\prime \prime }}{3+2\omega
_{{\rm BD}}^{0}}\,\,\,\,,
\end{eqnarray}
and therefore 
\begin{equation}
\Phi _{N}=-\frac{G_{0}M}{F_{0}r}\left[ 1+\frac{\sigma e^{-m_{0}r}}{3+2\omega _{%
{\rm BD}}^{0}}\right] \,\,\,\,.  \label{Newtpot2}
\end{equation}

Recently experimental bounds on the strength and range of a Yukawa
potential that could arise from ``fifth force fields'' have been analyzed 
in two kinds of experiments. The first kind consists in the analysis 
of gravitational signals induced by variations on the mass of a
lake \cite{Baldi}. Such experiments probe basically fields with a range 
$\lambda $ from meters to some kilometers and strength $|\beta |\in
[10^{-4},10^{2}]$. Here the coefficients corresponds to $\beta
=\sigma \,e^{-r/\lambda }/(3+2\omega _{{\rm BD}}^{0})$ and $\lambda =1/m_{0}$ (cf.
the curve $\pm \beta =\pm \beta (\lambda )$ in Ref.\cite{Baldi}) . 
The second kind of experiments probes variations of the Newton's 
gravitation law at scales of two Earth's radii by measuring the 
gravitational effects on the orbit of the laser-ranged LAGEOS satellite 
\cite{iorio}.  Assuming that 
$r\ll \lambda$, so that effects of an ``intermediate-range'' force are taken only 
to order $(r/\lambda)^2$, it turns that $|\beta | < 10^{-5}-10^{-8}$. These 
bounds are even more restrictive than those from the Earth based experiments 
quoted above.

At solar-system scales, Viking-like experiments restrict $\omega _{{\rm BD}
}^{0}>\omega _{{\rm \exp }},$ where $\omega _{{\rm \exp }}\sim 3000$
corresponds to the current lower bound on $\omega _{{\rm BD}}^{0}$ 
\cite{will0}. This bound results from the parametrized post-Newtonian(PPN) approximation and translates into a bound on $\beta $ and $\lambda$:
\begin{eqnarray}
\beta &\lesssim &\frac{\sigma\,e^{-r/\lambda }}{(3+2\omega _{{\rm \exp }})}\sim
1.66\,\,\,10^{-4}\sigma\,\,e^{-r/\lambda }\,\,. \\
\lambda &\lesssim &\sqrt{\frac{3+2\omega _{{\rm \exp }}}{2\omega _{{\rm \exp 
}}V_{0}^{\prime \prime }}}\sim 1/\sqrt{V_{0}^{\prime \prime }}.
\end{eqnarray}
If we assume for instance that $V=m^{2}\phi ^{2}$ and $m_{0}\ll $ eV , for
example, $m_{0}\sim 10^{-21}$ eV, then it turns that $\lambda \lesssim 13$
kpc, and moreover from Eq. (\ref{AA}) one can expect that $\sigma\sim 1$. Then,
the most stringent upper bound  $\sim 1/(3+2\omega _{{\rm BD}}^{0})$, 
imposed on $\beta \,$by the solar system experiments arises for long range
scalar fields $\lambda $ $\sim $ kpc (i.e., galactic-scale ranges) :

\begin{eqnarray}
\beta  &\lesssim &1.66\,\,\,10^{-4}\,\,.  \label{betacond2} \\
r & \ll & \,{\rm kpc}\,\,\,\,.
\end{eqnarray}

However, at the weak-field limit, a violation of the condition  (\ref
{betacond2}) do not imply significative violations of the $1/r^{2}$
gravitational force but rather a renormalization of the gravitational
constant or of the self-gravitating mass (see below). At the PPN level, the
effects of such ``massless'' scalar fields manifest sensitively in the propagation of
electromagnetic radiation (radar-echo delay).

Conversely, the solar system experiments impose weak bounds on scalar-tensor
theories if the scalar field has short range, since then the Yukawa
interaction would not be felt out of the self-gravitating body. However, as
pointed out above, at regimes of Earth range the ``fifth force tests'' start 
playing some role and therefore deviations of the $1/r^{2}$ gravitational
force could arise. Moreover, if $\lambda $ $\sim $ kpc, then also at
galactic scales a violation of the $1/r^{2}$ law could be expected.

{\bf Effective gravitational constant and total gravitational mass}. If we
assume for instance that $V=m^{2}\phi ^{2}$ then the self-consistency at
first order according to the conditions (\ref{V0cond}), implies that $m=0$
or $\phi _{0}=0$ (or both). In order to facilitate the interpretation of the
following considerations, let us assume that $V=0$ which corresponds to a
massless scalar field.

Now, the usual interpretation of fundamental constants in physics is by
identifying the coefficients of some fundamental fields in the corresponding
Lagrangian. For instance, one identifies the electric charge from the
coefficient of $1/e^{2}F_{\mu \nu }F^{\mu \nu }$ in the electromagnetic
Lagrangian, or the mass of fermions as the coefficient of $m\overline{\psi }
\psi $ in the Dirac Lagrangian. In the same way, the effective gravitational
constant can be identified as the coefficient of the Ricci scalar in the
Lagrangian (\ref{jordan}), that is by Eq. (\ref{Geff}). This means that at
zero order \cite{comm3},
\begin{equation}
G_{{\rm eff}}^0:=\frac{G_{0}}{F_{0}}\,\,\,\,.  \label{Geff0}
\end{equation}
So rather to define an effective gravitational constant from Eq. (\ref
{Newtpot2}): 
\begin{equation}
G:=\frac{2G_{0}}{F_{0}}\left( \frac{2+\omega _{{\rm BD}}^{0}}{3+2\omega _{%
{\rm BD}}^{0}}\right) \,\,\,\,,
\end{equation}
as it is usual in most treatments that have analyzed the weak field limit,
we shall keep the definition of Eq. (\ref{Geff0}). Instead one can define an
effective mass 
\begin{eqnarray}
M_{{\rm eff}}&:=& M\left[ 1+\frac{\alpha }{2G_{0}}\frac{(F_{0}^{\prime })^{2}}{
F_{0}}\right] =2M\left( \frac{2+\omega _{{\rm BD}}^{0}}{3+2\omega _{{\rm BD}
}^{0}}\right)\nonumber \\ 
&=& M+\frac{F_{0}^{\prime }Q}{2G_{0}}\,\,\,\,.
\end{eqnarray}
At this regard one mentions that it is the combination $G_{{\rm eff}}M_{{\rm eff}}$ 
that always appear in the Newtonian potential (\ref{Newtpot2}) and
not the separate quantities. So whether one adopts one or the other
definition for $G_{{\rm eff}}$ or $M_{{\rm eff}}$, the combination of both 
provides the same number.

Naively one could then expect that $M_{{\rm eff}}$ is the total
gravitational mass of the system. However, this is not the case, since the
total ADM-mass formula in the weak-field limit gives [see Eq.(\ref{Madmlin}) below]
\begin{equation}
M_{\rm ADM}=\int_{\Sigma _{t}}\tilde{T}^{00}d^{3}x\,\,\,\,,  \label{Madm}
\end{equation}
which turns to be 
\begin{eqnarray}
M_{{\rm ADM}}&=& M\left[ 1-\frac{\alpha }{2G_{0}}\frac{(F_{0}^{\prime })^{2}}{
F_{0}}\right] =2M\left( \frac{1+\omega _{{\rm BD}}^{0}}{3+2\omega _{{\rm BD}
}^{0}}\right) \nonumber \\
&=& M-\frac{F_{0}^{\prime }Q}{2G_{0}}\,\,\,\,.  \label{Madm2}
\end{eqnarray}
This result is not surprising if one takes into account that it is in fact 
the {\it active mass} $M_{{\rm eff}}$ and not the ADM-mass which is measured
by orbiting test particles \cite{will}. 

It is customary in the literature of gravitational physics to introduce
different mass definitions of a body. For instance, {\it tensor}, {\it scalar%
} and {\it inertial} (or sometimes referred to also as {\it Keplerian})
masses $M_{T}$, $M_{S}$ and $M_{I}$ \cite{will,Zag}. 
In the weak-field limit these are given
by 
\begin{eqnarray}
M_{T}\approx &&M\,\,\,\,, \\
\label{scmass}
M_{S}\approx &&\frac{M}{3+2\omega _{{\rm BD}}^{0}}= 
\frac{F_{0}^{\prime }Q}{2G_{0}}\,\,\,\,, \\
M_{I} \approx &&M_{{\rm eff}}\,\,\,\,.
\end{eqnarray}
In this limit, the tensor mass corresponds to the total rest mass since $M$
is just the integral of the energy-density of the self-gravitating body
which at first order coincides with the integral of the rest-mass density over the
proper-volume \cite{comm2} . The scalar mass (\ref{scmass}) 
is simply proportional to the scalar charge of the body. 
 The ADM-mass then reads in this
limit as, 
\begin{equation}
M_{{\rm ADM}}=M_{T}-M_{S}=M_{I}-2M_{S}=2M_{T}-M_{I}\,\,\,\,,
\end{equation}
and the other masses can be expressed in terms of the following
combinations: 
\begin{eqnarray}
M_{T} &=&M_{{\rm ADM}}+M_{S}=M_{I}-M_{S}\,\,\,\,, \\
M_{I} &=&M_{{\rm ADM}}+2M_{S}=M_{T}+M_{S}\,\,\,\,, \\
M_{S} &=&M_{{\rm ADM}}+2M_{S}=\frac{1}{2}\left( M_{I}-M_{{\rm ADM}}\right)
\,\,\,\,.
\end{eqnarray}
 \bigskip

{\bf Spontaneous scalarization.} The phenomenon of spontaneous scalarization 
\cite{damour3,damour2,novak,salgado2,whinnett} arises in scalar-tensor theories within a
dense compact object (e.g. neutron star). This phenomenon corresponds to a
``sudden'' appearance of a non-trivial configuration of the scalar field
when the extended object turns to be compact enough and in the absence of
scalar-field sources (as opposed to {\it induced scalarization}). In other
words, the spontaneous scalarization phenomenon in neutron stars corresponds
to a solution of field equations with fixed total baryon number and a zero
value for the asymptotic boundary condition $\phi _{0}$. In this way, there
are two possible configurations: one with a trivial scalar field ($\phi =0$)
with larger ADM-mass and another with a non-trivial scalar field with lower
ADM-mass, both configurations having the same total baryon number and $\phi
_{0}=0$. It has been argued that spontaneous scalarization is a
non-perturbative effect \cite{damour3} and therefore it is not expected to appear 
in the weak-field limit. This conclusion can be checked by analyzing the
regular solution for $\tilde{\phi}$ in Eq.(\ref{laplacephi}). In order to have 
spontaneous scalarization one would require $F_{0}^{\prime }= 0$ and at the 
same time $F_{0}^{\prime } \neq 0 $ (with $\phi_{0}=0$), both conditions 
needed to obtain a trivial and a regular non-trivial configurations for $\tilde \phi$. 
Since a multi-valued derivative is impossible for analytic functions 
$F(\phi)$, one concludes that spontaneous scalarization takes no place at the 
weak-field level. In other words, in the 
weak-field limit, the asymptotic 
condition $\phi_{0}=0$ fixes for one and for all the value of the constant 
$F_{0}^{\prime }$. This value has to be zero for the $\tilde \phi=0$ configuration to 
exist (the ADM-mass (\ref{Madm2}) has then the unique possible value $M_{{\rm ADM}}= M$) 
and different from zero to find the non-trivial 
scalarization solution $\tilde \phi\neq 0$ (both conditions are then impossible). 
However, in the non-perturbative 
analysis and still in the massless case $V=0$, one concludes from Eq. (\ref{KG})  
that a necessary condition for spontaneous scalarization to ensue is that 
$F$ has a critical point in $\phi=0$ (i.e., $\partial_\phi F|_{\phi=0}) =0$, 
with $T_{\rm matt}\neq 0$), since only in that case 
the solution $\phi=0$ is possible. For instance, an even function $F(\phi )$ verifies that
condition. The other necessary and sufficient condition to 
obtain spontaneous scalarization in the full theory results by proving that there exists a 
non-trivial stationary regular solution of the non-linear Klein-Gordon equation Eq. (\ref{KG}) 
for a configuration leaving the rest-mass unchanged and for the same asymptotic condition 
$\phi_{0}=0$. 
This has been done only numerically and for some class of STT \cite{damour2,damour3,novak,salgado2}. 
For example, the particular case $F=1+16\pi \xi \phi ^{2}$, was analyzed non
perturbatively in Ref. \cite{salgado2} via a numerical analysis and the 
conclusion was that only within highly compact neutrons stars (strong
field regime) the phenomenon appeared.

As opposed to spontaneous-scalarization there is the {\it induced
scalarization}. This arises when $\phi _{0}\neq 0$, and since any analytic (non-trivial) 
function $F(\phi) $ implies $F_{0}^{\prime }\neq 0$ for $\phi _{0}\neq 0$, 
it has a weak-field limit analogue (it would be rather artificial to expect 
a Lagrangian which depends explicitly on $\phi _{0}$ and moreover that 
$F_{0}^{\prime }=0$). The simplest non-trivial example is $F=1+16\pi \xi
\phi $. In this case $F^{\prime }|_{\phi _{0}}=const.$ and therefore a
trivial solution for $\tilde{\phi}$ is not allowed [cf. Eq.(\ref{KG}) or (
\ref{laplacephi}) with $V=0$]. An artificial coupling $F=1+16\pi \xi (\phi
-\phi _{0})^{2}$ leads however to $F^{\prime }|_{\phi _{0}}=0$. In other
words, with the asymptotic condition $\phi _{0}\neq 0$ and with a coupling
function $F(\phi ,\phi _{0})$ such that $\partial _{\phi _{0}}F=0$, it is
impossible to have a trivial scalar field in any limit (weak or strong
regime), since a trivial scalar-field $\phi =const.$ would not be a solution
of Eq. (\ref{KG}) for $T_{{\rm matt}}\neq 0$. Induced scalarization then
always ensues. The fact that the induced scalarization is present in the
weak-field limit produces a smooth transition to the strong field regime (as
opposed to spontaneous scalarization). Namely, the curve $Qvs.M_{{\rm bar}}$
(scalar charge vs. total baryon mass) do not show any discontinuity contrary 
to spontaneous scalarization (cf. Ref. \cite{damour2,damour3,salgado2}).

\section{Conclusions}
The linear limit of scalar-tensor theories of gravity leads to the prediction of 
a spin-zero gravitational mode (scalar waves) in addition to the well known spin-2 
modes. In the weak-field limit, STT predicts an effective gravitational potential 
with a Yukawa contribution for massive scalar fields. 
However, for ultra-light 
scalar fields (scalar fields of galactic or cosmological range) the Yukawa term 
is highly supressed as far as solar system experiments are 
concerned. Thus, in such a case the effective gravitational potential leads 
essentially to a Newtonian gravitational force $\sim 1/r^2$ with an effective mass 
coefficient which is proportional to the rest-mass of the self-gravitating body 
and which depends explicity on the 
effective Brans-Dicke parameter. This latter is highly constrained by the 
post-Newtonian approximation via Viking-like experiments. Therefore, even if a
 STT does not produce a deviation of the $1/r^2$ Newton's law, it can however 
violate dramatically the post-Newtonian bounds. On the other hand, if the 
scalar field has ``intermediate'' ranges (meters to kilometers or even 
thousands of kilometers), then the Yukawa term leads to 
a ``fifth force'' field that produces deviations of the $1/r^2$ 
Newton's law and which are severely constrained by Earth based experiments 
or by satellites, notably, by varying-mass experiments and laser-ranged missions. 
STT also predicts the phenomenon of 
spontaneous scalarization in compact objects. For the phenomenon to take 
place, a STT requires the existence of 
two possible configurations, one of which corresponds to the trivial 
$\phi=0$ solution and another one that is not trivial. Both configurations 
having the same rest-mass and the null asymptotic value $\phi_0=0$. 
The phenomenon will not be present in the weak-field limit since it would 
require that the non-minimal 
coupling function $F(\phi)$ has a two valued-derivatives: $F'(\phi)|_{\phi_0=0} =0$ and 
$F'(\phi)|_{\phi_0 = 0}\neq 0$ in order two obtain the two possible configurations for 
the perturbation $\tilde \phi$. The conclusion is that no-analytic function $F(\phi)$ 
can fullfill both conditions.  As oppposed to spontaneous scalarization, 
induced scalarization takes place also in the weak-field limit, since 
 for physical couplings one has $F'(\phi_0) \neq 0$  (i.e. physical 
coupling functions $F$ will not depend explicitly on $\phi_0$), and then only non-trivial regular 
solutions are admitted.

\begin{acknowledgments}
I'm indebted to O.Obreg\'on, U. Nucamendi and D. Sudarsky for fruitful discussions. 
This work has been supported by DGAPA-UNAM, grant Nos. IN112401 and 
IN119799 and CONACYT, M\'exico, grant No. 32551-E.
\end{acknowledgments}

\appendix

\section{The 3+1 formulation of general relativity}
Let us consider the 3+1 or Adison-Deser-Misner (ADM) formulation of general 
relativity in which the spacetime (considered to be globally hyperbolic) 
is foliated by a family of spacelike hypersurfaces $\Sigma_t$. 
We shall not enter into the details of the derivation of the 3+1 equations (see
Refs. \cite{ADM,Choquet,msnotes,york79,wald}). The sign convention for the 
3+1 splitting of the metric is as follows:
\begin{equation}  \label{3+1metric}
ds^2= -(N^2 - N^iN_i)dt^2 - 2N_i dt dx^i + h_{ij} dx^i dx^j\,\,\,.
\end{equation}
The extrinsic curvature of the embedings $\Sigma_t$ is given by
\begin{eqnarray}  \label{K_ij}
K_{ij} &=& -\nabla_i n_j= -N\Gamma^t_{ij} \nonumber \\
&=& -\frac{1}{2N}\left( \frac{\partial
h_{ij}}{\partial t} + \,^3\nabla_j N_i + \,^3\nabla_i N_j \right) \,\,\,,
\end{eqnarray}
where $\,^3\nabla_j$ stands for the covariant derivative compatible with the 
three-metric $h_{ij}$. This is to be regarded as an evolution equation for 
$h_{ij}$. The trace of the extrinsic curvature will be denoted by, 
\begin{equation}  \label{divK2}
K:= K^{l}_{\,\,\,l} \,\,\,\,.
\end{equation}

The orthogonal decomposition of the energy-momentum tensor in components
tangent and orthogonal to $\Sigma_t$ leads to \cite{york79}:

\begin{equation}  \label{Tort}
T^{\mu \nu} = S^{\mu \nu} + J^{\mu} n^{\nu} + n^\mu J^{\nu} + E n^\mu
n^\nu\,\,\,\,.
\end{equation}
where $n^\mu$ is the normal to $\Sigma_t$. 
The tensor $S^{\mu \nu}$ is symmetric and often called the {\it tensor of
constraints}; $J^\mu$ is the {\it momentum density vector} and $E$ is the 
{\it total energy density} measured by the observer 
orthogonal to $\Sigma_t$. 

As in the 4+0 formalism, $T^{\mu\nu}$
will be the total energy-momentum tensor of matter which can be composed by
the contribution of different types of sources: 
\begin{equation}  \label{Ttot}
T^{\mu\nu}= \sum_i T^{\mu \nu}_i\,\,\,\,.
\end{equation}
This means that 
\begin{equation}  \label{3+1mattvartot}
E = \sum_i E_i \,\,\,,\,\,\, J^\mu = \sum_i J^\mu _i \,\,\,,\,\,\,
S^{\mu\nu} = \sum_i S^{\mu\nu}_i \,\,\,\,.
\end{equation}

The projection of Einstein equations $R_{\mu\nu}= 4\pi G_0\left(2T_{\mu\nu}-
T^{\alpha}_{\,\,\,\alpha} g_{\mu\nu}\right)$ in the directions tangent and
orthogonal to $\Sigma_t$, followed by the use of the Gauss-Codazzi-Mainardi
equations leads to the 3+1 form of Einstein equations:

\begin{equation}  \label{CEHf}
^3 R + K^2 - K_{ij} K^{ij}= 16\pi G_0 E \,\,\,,
\end{equation}
known as the Hamiltonian constraint.

\begin{equation}  \label{CEMf}
^3 \nabla_l K_{\,\,\,\,\,i}^{l} - \,^3\nabla_i K= 8\pi G_0 J_i \,\,\,\,,
\end{equation}
known as the {\it momentum constraint} equations.

Finally, the {\it dynamic} Einstein equations read 
\begin{eqnarray}  \label{EDEf}
& & \partial_t K_{\,\,\,j}^i + N^l \partial_l K_{\,\,\,j}^i + K_{\,\,\,l}^i
\partial_j N^l - K_{\,\,\,j}^l \partial_l N^i \nonumber \\ 
&& + \,^3\nabla^i\,^3\nabla_j N  -\,^3 R_{\,\,\,j}^i N - N K K_{\,\,\,j}^i  \nonumber \\
&& = 4 \pi G_0 N\left[ (S-E)\delta^i_{\,\,j} - 2S^i_{\,\,j}\right]\,\,\,\,
\end{eqnarray}
where $S= S^l_{\,\,l}$ is the trace of the tensor of constraints, and all
the quantities written with a `$3$' index refer to those computed with the
three-metric $h_{ij}$. Moreover, under the 3+1 formalism tensor quantities
tangent to $\Sigma_t$ use the three-metric to raise and lower their spatial
indices. The equations (\ref{K_ij}) and (\ref{EDEf}) are the set of the
Cauchy-initial-data evolution equations for the gravitational field subject
to the constraints Eqs. (\ref{CEHf}) and (\ref{CEMf}).

We can write an evolution equation for the trace $K$ by taking the trace in
Eq. (\ref{EDEf}): 
\begin{equation}  \label{EDK}
\partial_t K + N^l \partial_l K + \,^3\Delta N -N\left( \,^3 R + K^2\right)
= 4 \pi G_0 N\left[S-3 E\right] \,\,\,.
\end{equation}
where $\,^3\Delta$ stands for the Laplacian operator compatible with $h_{ij}$%
.

This can be simplified by using Eq.(\ref{CEHf}) to give 
\begin{equation}  \label{EDK2}
\partial_t K + N^l \partial_l K + \,^3\Delta N -N K_{ij} K^{ij} = 4 \pi G_0
N\left[S + E\right] \,\,\,.
\end{equation}

For our purposes it will no be necessary to write the 3+1 equations for the
matter and the scalar field.

\subsection{Linearized equations}

In order to linearize the 3+1 equations we assume first order deviations of
the 3+1 metric with respect to the Lorentz metric $\eta _{\mu \nu }$ as
follows 
\begin{eqnarray}
N &\approx &1+\epsilon \tilde{N}\,\,\,\,\,, \\
N_{i} &\approx &\epsilon \tilde{N}_{i}\,\,\,\,\,, \\
h_{ij} &\approx &\delta _{ij}+\epsilon \tilde{h}_{ij}\,\,\,\,\,.
\end{eqnarray}
At first order $h^{ij}\approx \delta _{ij}-\epsilon \tilde{h}_{ij}$ in order
to satisfy the inverse condition $h_{il}h^{lj}=\delta _{\,\,\,j}^{i}$.
Therefore, covariant and contravariant components of tensorial quantities
having no zero order terms are identical to each other at first order. For
instance $N^{i}=h^{il}N_{l}\approx \epsilon \tilde{N}_{i}$. In order to
compare with the 4+0 (full covariant) linear approximation 
\begin{equation}
g_{\mu \nu }\approx \eta _{\mu \nu }+\epsilon \gamma _{\mu \nu }\,\,\,\,,
\end{equation}
it turns out that 
\begin{eqnarray}
\gamma _{00} &=&-2\tilde{N}\,\,\,\,,  \label{Newtpot3p1} \\
\gamma _{0i} &=&-\tilde{N}_{i}\,\,\,\,, \\
\gamma _{ij} &=&\tilde{h}_{ij}\,\,\,\,.
\end{eqnarray}

The 3-Christoffel symbols turn to be 
\begin{eqnarray}
^{3}\,\!\Gamma _{jk}^{i} &\approx &\epsilon \,\,^{3}\,\!\tilde{\Gamma}%
_{jk}^{i}\,\,\,\,, \\
^{3}\,\!\tilde{\Gamma}_{ij}^{i} &:=&\frac{1}{2}\left( -\partial _{i}\tilde{h}%
_{jk}+\partial _{k}\tilde{h}_{ij}+\partial _{j}\tilde{h}_{ki}\right)
\,\,\,\,.
\end{eqnarray}
Therefore 3-covariant derivatives of 3-tensors having no zero order terms
become at first order ordinary derivatives. For instance $^{3}\nabla
_{j}N_{i}\approx \epsilon \partial _{j}\tilde{N}_{i}$. Then Eq.(\ref{K_ij})
leads to 
\begin{eqnarray}
K_{ij} &\approx &\epsilon \tilde{K}_{ij}\,\,\,\,, \\
\label{linK_ij}
\tilde{K}_{ij} &:= &-\frac{1}{2}\left( \partial _{t}\tilde{h}_{ij}+2\partial
_{(i}\tilde{N}_{j)}\right) \,\,\,\,,
\end{eqnarray}
with 
\begin{eqnarray}
K_{\,\,\,j}^{i} &\approx &\epsilon \tilde{K}_{\,\,\,j}^{i}\approx \epsilon 
\tilde{K}_{ij}\,\,\,\,,\\
K &:= &K_{\,\,\,l}^{l}\approx \epsilon \tilde{K}\,\,\,\,, \\
\label{Klin}
\tilde{K} &:=& -\frac{1}{2}\left( \partial _{t}\tilde{h}+2\partial _{l}\tilde{%
N}_{l}\right) \,\,\,\,,
\end{eqnarray}
where $\tilde{h}:=\tilde{h}_{ll}$ is the trace of the 3-metric perturbation.
Using the definition of the 3-Riemann tensor in terms of the 3-Christoffel
symbols it is easy to obtain the linearized approximation for the 3-Ricci
tensor and the 3-curvature respectively: 
\begin{eqnarray}
^{3}R_{ij} &\approx &\epsilon \,\,^{3}\tilde{R}_{ij}\,\,\,\,, \\
^{3}\tilde{R}_{ij} &:= &\frac{1}{2}\left( 
2\partial_{l(j}^{2}\tilde{h}_{i)l}^{}-\,^{3}\tilde{\Delta}\tilde{h}_{ij}-\partial _{ij}^{2}\tilde{h}\right)
\,\,\,\,, \\
^{3}R &\approx &\epsilon \,\,^{3}\tilde{R}\,\,\,\,, \\
^{3}\tilde{R} &:= &\partial _{kl}^{2}\tilde{h}_{kl}-\,^{3}\tilde{\Delta}%
\tilde{h}\,\,\,\,,
\end{eqnarray}
where $^{3}\tilde{\Delta}:=\partial _{ll}^{2}$ stands for the Euclidean
3-Laplacian.

Concerning the sources (matter fields) we assume 
\begin{eqnarray}
E&\approx & E_0 + \epsilon \tilde E \,\,\,\,, \\
J_i &\approx & J^0_i + \epsilon \tilde J_i \,\,\,\,, \\
S_{ij} &\approx & S_{ij}^0 + \epsilon \tilde S_{ij} \,\,\,\,.
\end{eqnarray}

The Hamiltonian constraint Eq. (\ref{CEHf}) when linearized reads 
\begin{equation}  \label{CEHflin0}
^3 \tilde R = 16\pi G_0 \tilde E \,\,\,,
\end{equation}
or explicitly 
\begin{equation}  \label{CEHflin}
\partial^2_{kl}\tilde h_{kl} - \,^3\tilde \Delta \tilde h = 16\pi G_0 \tilde %
E \,\,\,.
\end{equation}

The linearized version of the momentum constraint Eq. (\ref{CEMf}),  reads 
\begin{equation}  \label{CEMflin}
\partial_l \tilde K_{\,\,\,\,\,i}^{l} - \,\partial_i \tilde K = 8\pi G_0 
\tilde J_i \,\,\,\,.
\end{equation}

Finally the dynamic Einstein Eqs. (\ref{EDEf}) when linearized read 
\begin{equation}  \label{EDEflin}
\partial_t \tilde K_{\,\,\,j}^i + \partial^i\,\partial_j \tilde N - \,^3 
\tilde R_{\,\,\,j}^i = 4 \pi G_0 \left[ (\tilde S-\tilde E)\delta^i_{\,\,j}
- 2\tilde S^i_{\,\,j}\right]\,\,\,\,.
\end{equation}

The linear limit of evolution Eq. (\ref{EDK2}) reads 
\begin{equation}  \label{EDK2lin}
\partial_t \tilde K + \,^3\tilde \Delta \tilde N = 4 \pi G_0 \left[\tilde S
+ \tilde E\right] \,\,\,.
\end{equation}

The self-consistency of the 3+1 equations up to first order imply that the
zero order source fields must vanish identically: 
\begin{eqnarray}
E_0 = 0 = J^0_i = S_{ij}^0 \,\,\,\,.
\end{eqnarray}

The above equations (\ref{CEHflin}), (\ref{CEMflin}) and (\ref{EDEflin})
are the 3+1 decomposition of the 4+0 equations (\ref{einstlin}).

The 3+1 splitting of the perturbations (\ref{tilgamma}) is 
\begin{eqnarray}
\tilde{\gamma}_{00} &=&\frac{1}{2}\left( \tilde{h}-2\kappa \tilde{\phi}-2%
\tilde{N}\right) \,\,\,\,, \\
\tilde{\gamma}_{0i} &=&-\tilde{N}_{i}\,\,\,\,, \\
 \label{tilgammaij}
\tilde{\gamma}_{ij} &=&\tilde{h}_{ij}-\frac{1}{2}\delta _{ij}\left( \tilde{h}%
-2\kappa \tilde{\phi}+2\tilde{N}\right) \,\,\,\,, \\
\tilde{\gamma} &=&-\left( \tilde{h}+2\tilde{N}\right) +4\kappa \tilde{\phi}%
\,\,\,\,,
\end{eqnarray}

The 3+1 splitting of the Lorentz gauge Eq. (\ref{radgaugescal0}) 
\begin{eqnarray}
&& \partial ^{\mu }\tilde{\gamma}_{\mu \nu }= 0 \nonumber \\
&&=\,\,\left\{\begin{array}{c} 
\partial_t \tilde h -2 \kappa \partial_t\tilde \phi + 2 \partial_i \tilde
N^i - 2 \partial_t \tilde N = 0  \,\,\, {\rm for}\,\,\,\nu=0 \\
\partial_t \tilde N^i + \partial_j \tilde h_{ij} -\partial_i \tilde N
-\frac{1}{2}\partial_i\tilde h + \kappa \partial_i \tilde \phi = 0  \,\,\,
{\rm for}\,\,\,\nu=i 
\end{array}\right. \,\,\,\, \nonumber \\
&&  \label{radgaugescal}
\end{eqnarray}
Using Eq. (\ref{Klin}) the zero component of Eq. (\ref{radgaugescal}) reads 
\begin{equation}
\partial _{t}\tilde{N}+\kappa \partial _{t}\tilde{\phi}=-\tilde{K}\,\,\,\,.
\label{radgauge0scal}
\end{equation}
From Eqs. (\ref{radgauge0scal}) and (\ref{EDK2lin}) one obtains a wave equation for $\tilde{N}$: 
\begin{equation}
-\partial _{tt}^{2}\tilde{N}+\,^{3}\tilde{\Delta}\tilde{N}=\kappa \partial
_{tt}^{2}\tilde{\phi}+4\pi G_{0}\left[ \tilde{S}+\tilde{E}\right] \,\,\,.
\label{waveNscal}
\end{equation}
On the other hand, differentiating the spatial components of the Lorentz
gauge Eq. (\ref{radgaugescal}) with respect to time and using consecutively
Eqs. (\ref{linK_ij}), (\ref{Klin}) and the momentum constraint Eq. (\ref
{CEMflin}) one obtains a wave equation for $\tilde{N}^{i}$: 
\begin{equation}
-\partial _{tt}^{2}\tilde{N}^{i}+\,^{3}\tilde{\Delta}\tilde{N}^{i}=2\kappa
\partial _{ti}^{2}\tilde{\phi}-16\pi G_{0}\tilde{J}^{i}\,\,\,.
\label{waveNiscal}
\end{equation}
Finally, the linearized dynamic Einstein Eqs. (\ref{EDEflin}) together with
Eq. (\ref{linK_ij}) and the spatial components of the Lorentz gauge Eq. (\ref
{radgaugescal}) lead to a wave equation for $\tilde{h}_{ij}$ 
\begin{equation}
-\partial _{tt}^{2}\tilde{h}_{ij}+\,^{3}\tilde{\Delta}\tilde{h}%
_{ij}=-2\kappa \partial _{ij}^{2}\tilde{\phi}+8\pi G_{0}\left[ (\tilde{S}-%
\tilde{E})\delta _{\,\,j}^{i}-2\tilde{S}_{\,\,j}^{i}\right] \,\,\,\,.
\label{wavehijscal}
\end{equation}
From Eq. (\ref{wavehijscal}) one obtains 
\begin{equation}
-\partial _{tt}^{2}\tilde{h}+\,^{3}\tilde{\Delta}\tilde{h}=-2\kappa \,^{3}%
\tilde{\Delta}\tilde{\phi}+8\pi G_{0}\left[ \tilde{S}-3\tilde{E}\right]
\,\,\,\,.  \label{wavehscal}
\end{equation}
Another wave equation which is not independent from the above is obtained
from the linearized Hamiltonian constraint Eq. (\ref{CEHflin}) and the 
Lorentz-gauge conditions Eq. (\ref{radgaugescal}): 
\begin{equation}
\left( -\partial _{tt}^{2}+\,^{3}\tilde{\Delta}\right) \,\left( \frac{1}{2}%
\tilde{h}-\tilde{N}\right) =-\kappa \left( \partial _{tt}^{2}+\,^{3}\tilde{%
\Delta}\right) \tilde{\phi}-16\pi G_{0}\tilde{E}\,\,\,\,.
\end{equation}
which is obtained by combining Eqs. (\ref{waveNscal}) and (\ref{wavehscal}). This
can be written as 
\begin{equation}
\left( -\partial _{tt}^{2}+\,^{3}\tilde{\Delta}\right) \left( \frac{1}{2}%
\tilde{h}-\kappa \tilde{\phi}-\tilde{N}\right) =-2\kappa \,^{3}\tilde{\Delta}%
\tilde{\phi}-16\pi G_{0}\tilde{E}\,\,\,\,.  \label{wavebargamma00scal}
\end{equation}

The combination of Eqs. (\ref{waveNscal}), (\ref{wavehijscal}) and (\ref
{wavehscal}) as given by Eq. (\ref{tilgammaij}) provides the wave equation 
\begin{equation}  \label{wavetilgammaij}
\left(-\partial^2_{tt} + \,^3\tilde \Delta\right)\tilde \gamma_{ij}=
-2\kappa \partial^2_{ij}\tilde \phi + 2\kappa \delta_{ij}\Box_\eta \tilde\phi
-16\pi G_0 \tilde S_{ij} \,\,\,\,.
\end{equation}
Therefore Eqs. (\ref{waveNiscal}), (\ref{wavebargamma00scal}) and (\ref
{wavetilgammaij}) recover the 4+0 wave Eq. (\ref{wavetilgamma}) in the Lorentz gauge.

Now according to the 3+1 splitting of the energy-momentum tensor, and using
Eq. (\ref{Tmunulin}) we have 
\begin{eqnarray}
\tilde{E} &=&\tilde{T}_{00}=\frac{\tilde{E}_{{\rm matt}}}{F_{0}}+\frac{
F_{0}^{\prime }}{8\pi F_{0}G_{0}}\,^{3}\tilde{\Delta}\tilde{\phi}\,\,\,\,,
\label{Eeuler} \\
\tilde{J}_{i} &=&-\tilde{T}_{i0}=-\frac{\tilde{T}_{0i}^{{\rm matt}}}{F_{0}}-
\frac{F_{0}^{\prime }}{8\pi F_{0}G_{0}}\partial _{0i}^{2}\tilde{\phi}
\,\,\,\,, \\
\tilde{S}_{ij} &=&\tilde{T}_{ij}=\frac{\tilde{T}_{ij}^{{\rm matt}}}{F_{0}}+
\frac{F_{0}^{\prime }}{8\pi F_{0}G_{0}}\left( \partial _{ij}^{2}\tilde{\phi}
-\delta _{ij}\Box _{\eta }\tilde{\phi}\right) \,\,\,\, \\
\tilde{S} &=&\tilde{T}_{\,\,\,l}^{l}=\frac{\tilde{T}_{\,\,\,\,\,l}^{{\rm matt
}\,\,l}}{F_{0}}+\frac{F_{0}^{\prime }}{8\pi F_{0}G_{0}}\left( 3\partial
_{tt}^{2}\tilde{\phi}-2\,^{3}\tilde{\Delta}\tilde{\phi}\right) \,\,\,.\nonumber \\
&&
\end{eqnarray}
Then Eqs. (\ref{waveNscal}), (\ref{waveNiscal}), and (\ref{wavehijscal})
read respectively 
\begin{eqnarray}
-\partial _{tt}^{2}\tilde{N} &+& \,^{3}\tilde{\Delta}\tilde{N} = 4\pi \frac{G_{0}}{
F_{0}}\left( \tilde{E}^{{\rm matt}}+\tilde{S}^{{\rm matt}}\right) \nonumber \\
& + &\left(
\kappa +\frac{3F_{0}^{\prime }}{2F_{0}}\right) \partial _{tt}^{2}\tilde{\phi}
-\frac{F_{0}^{\prime }}{2F_{0}}\,^{3}\tilde{\Delta}\tilde{\phi}\,\,\,\,.
\label{waveNscal2}
\end{eqnarray}

\begin{equation}  \label{waveNiscal2}
-\partial^2_{tt}\tilde N^i + \,^3\tilde \Delta \tilde N^i = -16 \pi \frac{G_0
}{F_0}\tilde J^{i}_{{\rm matt}} + 2\left( \kappa + \frac{F^{\prime}_0}{F_0}
\right) \partial^2_{ti}\tilde\phi \,\,\,\,.
\end{equation}

\begin{eqnarray}  \label{wavehijscal2}
-\partial^2_{tt}\tilde h_{ij} &+& \,^3\tilde \Delta \tilde h_{ij} = 8 \pi 
\frac{G_0}{F_0} \left[ (\tilde S_{{\rm matt}}-\tilde E_{{\rm matt}
})\delta^i_{\,\,j} - 2\tilde S^{{\rm matt}}_{ij}\right] \nonumber \\
&-& \frac{F^{\prime}_0
}{F_0}\delta^i_{\,\,j} \Box_\eta \tilde\phi - 2 \left(\kappa + \frac{
F^{\prime}_0}{F_0}\right) \partial^2_{ij}\tilde\phi \,\,\,\,.
\end{eqnarray}
From Eq. (\ref{wavehscal}) one obtains 
\begin{eqnarray}  
-\partial^2_{tt}\tilde h + \,^3\tilde \Delta \tilde h &=& 8 \pi \frac{G_0}{F_0}
\left[ \tilde S_{{\rm matt}} -3\tilde E_{{\rm matt}} \right] -3 \frac{
F^{\prime}_0}{F_0} \Box_\eta \tilde\phi \nonumber \\
&& - 2 \left(\kappa + \frac{F^{\prime}_0
}{F_0}\right) \,^3\tilde \Delta \tilde \phi \,\,\,\,. \label{wavehscal2}
\end{eqnarray}

Moreover, from Eq. (\ref{wavebargamma00scal}) we get 
\begin{eqnarray}
&& \left( -\partial _{tt}^{2}+\,^{3}\tilde{\Delta}\right) \left( \frac{1}{2}
\tilde{h}-\kappa \tilde{\phi}-\tilde{N}\right) = -16\pi \frac{G_{0}}{F_{0}}
\tilde{E}_{{\rm matt}}\nonumber \\
&& -2\left( \kappa +\frac{F_{0}^{\prime }}{F_{0}}\right)
\,^{3}\tilde{\Delta}\tilde{\phi}\,\,\,\,.  \label{wavebargamma00scal2}
\end{eqnarray}
Whereas the Eq. (\ref{wavetilgammaij}) leads to 
\begin{eqnarray}
&& \left( -\partial _{tt}^{2}+\,^{3}\tilde{\Delta}\right) \tilde{\gamma}
_{ij} = -16\pi \frac{G_{0}}{F_{0}}\tilde{S}_{ij}^{{\rm matt}} \nonumber \\
&& -2\left( \kappa +
\frac{F_{0}^{\prime }}{F_{0}}\right) \left( \partial _{ij}^{2}\tilde{\phi}
-\delta _{ij}\Box _{\eta }\tilde{\phi}\right) \,\,\,\,.
\label{wavetilgammaij2}
\end{eqnarray}
Therefore Eqs. (\ref{waveNiscal2}), (\ref{wavebargamma00scal2}), and (\ref
{wavetilgammaij2}) are equivalent to Eq. (\ref{wavetilgamma2}). In above
equations one can employ the convenient choice $\kappa =-F_{0}^{\prime
}/F_{0}$ used in the 4+0 formulation to simplify the expressions.


\subsection{The weak-field approximation}

\smallskip As in the 4+0 formulation, we consider slow varying sources and
neglect the time-derivatives, and also $\tilde{E}_{{\rm matt}}=\tilde{\rho}$
, $\tilde{S}^{{\rm matt}} \ll \tilde{\rho},\tilde{J}_{{\rm matt}}^{i} \ll \tilde{
\rho}c$. With such considerations Eq. (\ref{waveNscal2}) reads,

\begin{equation}
\,^{3}\tilde{\Delta}\left( \tilde{N}+\frac{F_{0}^{\prime }}{2F_{0}}\,\tilde{
\phi}\,\right) =4\pi \frac{G_{0}}{F_{0}}\tilde{\rho}\,.  \label{laplaceNscal}
\end{equation}

As in the 4+0 formulation $\tilde{\gamma}_{ij}={\rm const.}=\tilde{N}^{i}$
are the regular solutions and the constants can be gauged out. Combining 
(\ref{laplaceNscal}) and (\ref{laplacephi}) we can then obtain the solution 
given by Eq. (\ref{Newtpot}). Therefore the two
non-trivial degrees of freedom are $\tilde{N}$ and $\tilde{\phi}\,$( $\tilde{%
N}^{i}=0$, and, $h_{ij}$ is given in terms of $\tilde{N}$ and $\tilde{\phi}).
$ The perturbed quantity $\tilde{N}$ has the direct interpretation of the
Newtonian potential [cf. Eqs.(\ref{Newtpot0}) and (\ref{Newtpot3p1})]. In
particular, for the massless case $m_{0}=0$, it yields,

\begin{equation}
\,\tilde{\Delta}\tilde{N}=4\pi \frac{G_{0}\tilde{\rho}}{F_{0}}\left(
1+\alpha \frac{F_{0}^{\prime ^{2}}}{2G_{0}F_{0}}\right) \,,
\end{equation}

\smallskip where Eq. (\ref{laplacephi}) was used. In the spherical symmetric
case, (\ref{Newtpot2}) provides the exterior solution for $\tilde{N}.$

Concerning the mass issue, let us focus on the linearized Hamiltonian 
constraint Eq. (\ref{CEHflin}). One can integrate over a coordinate volume 
and use Gauss theorem to obtain
\begin{equation}\label{Madmlin}
\frac{1}{16\pi} \oint \left(\partial_{l}\tilde h_{kl} - \, \partial_k\tilde h\right)dS^k
= \int_{\Sigma _{t}} \tilde{E}d^{3}x \,\,\,\,,
\end{equation}
where the integrals extend to spatial infinite. 
The left-hand side corresponds precisely to the weak-field 
version of the ADM-mass formula [cf. Eq. (11.2.14) of Ref. \cite{wald}],
\begin{equation}\label{Madmex}
M_{\rm ADM}:= \frac{1}{16\pi} \oint \left(\partial_{l}h_{kl} - \, 
\partial_k h\right)dS^k \,\,\,\,,
\end{equation}
while the right-hand side of Eq. (\ref{Madmlin}) is the weak-field 
approximation of the Komar-mass formula.

In the massless case $m_{0}=0,$from (\ref{Eeuler}) and (\ref{laplacephi}),
it turns
\begin{equation}
\label{Eeulerlin}
\tilde{E}=\frac{\tilde{\rho}}{F_{0}}\left( 1-\frac{\alpha}{2G_{0}}
\frac{(F_{0}^{\prime})^2}{F_{0}}\right) \,\,\,.
\end{equation}

Therefore, from Eqs. (\ref{Madmlin}) and (\ref{Eeulerlin}), 
one recovers Eq.(\ref{Madm2}) for the ADM-mass in the weak-field
limit .


\begin{thebibliography}{}
\bibitem{comm}  However, we emphasize that the equation of hydrostatic
equilibrium is of second order since it involves terms $\tilde{\rho}\partial
_{r}\gamma _{00}$ $\sim $ $\partial _{r}p$ which provide the balance between
the gravitational force and the pressure gradients.
\bibitem{comm3}
At first order the effective gravitational constant, as given by Eq.(\ref{Geff}) is 
$G_{\rm eff}\approx (G_0/F_0) \left( 1- \epsilon \tilde \phi F'_0/F_0\right)$.
\end{thebibliography}
\end{document}